\documentclass[12pt,a4paper]{article}
\usepackage[utf8]{inputenc}
\usepackage[francais]{babel}
\usepackage[T1]{fontenc}
\usepackage{lmodern}
\usepackage{amsmath}
\usepackage{amsfonts}
\usepackage{amssymb}
\usepackage{setspace}

\usepackage{graphicx,}
\begin{document}

\title{Spectroscopie théorique des quasars et loi de Karlsson.}

\author{Jacques Moret-Bailly\footnote{email: jmo@laposte.net}}

\maketitle

\newpage

\begin{abstract}
La loi de Karlsson introduite par la spectroscopie des quasars peu rougis utilise le spectre Lyman de l'atome d'hydrogène. Il faudra comparer les concepts déduits de la spectroscopie classique des quasars, développée ici, à ceux qui sont déduits du $\Lambda$-CDM.
Une absorption fine et saturée d'une raie spectrale dans un gaz requiert un long parcours sans perturbation comme collisions ou rougissement cosmologique. Ainsi, les spectres de "forêts Lyman" des quasars obéissant à la loi de Karlsson résultent essentiellement d'interactions de la lumière thermique naturelle rayonnée par le quasar avec de l'hydrogène atomique froid sous très basse pression. Ces raies sont produites par trois processus: a) Une absorption usuelle produit un ensemble de raies; b) Ces raies sont multipliées par absorption après des rougissements fondamentaux 3K (ou 4K), où K est la constante de Karlsson; 3K (ou 4K) amènent exactement une raie Lyman beta (ou gamma) absorbée sur une raie Lyman alpha: La coincidence d'une raie absorbée rougie, avec Lyman alpha arrête donc quasiment le rougissement et les raies du gaz sont absorbées intensément; c) Les rougissements se produisent dans des régions où la lumière à la fréquence alpha est faiblement absorbée en raison du rougissement permanent, sauf lorsque l'excitation de l'hydrogène au niveau 2P est suffisante pour qu'un flash superradiant éclate à la fréquence alpha, ce qui provoque une absorption intense des rayons issus du quasar dont la radiance est grande, donc l'inscription d'une raie à la fréquence Lyman alpha courante. Éclairs et pompages produisent des  oscillations de relaxation qui inscrivent de nombreuses raies d'absorption. Les rougissements par les atomes H dans les niveaux 2P sont dus à des interactions paramétriques composées d'effets Raman impulsionnels stimulés (ISRS): Des atomes d'hydrogène excités catalysent des échanges d'énergie entre le rayon observé et des rayons froids du fond thermique, en accord avec la thermodynamique. La description de l'univers devient beaucoup plus simple mais moins merveilleuse.
\newpage

\centerline
\title{\bf Theoretical spectroscopy of quasars within Karlsson's law.}

\medskip
\centerline
{\bf Abstract.}

The law introduced by Karlsson in spectroscopy of low-redshift quasars involves the Lyman spectrum of hydrogen atoms. Thus, it appears necessary to study the concepts introduced by a standard spectroscopy of quasars, studied here, with those deducted from $\Lambda$-CDM.
A visible absorption of a sharp and saturated spectral line in a gas requires a long path without perturbations as collisions or cosmological redshift. Spectra of absorbed, saturated lines of quasars obeying Karlsson's law mainly result from interactions of natural, thermal light radiated by quasar with relatively cold, low presure atomic hydrogen. These lines are produced by three processes: a) A conventional absorption in a relatively cold gas produces a set of lines; b) These lines are multiplied by absorption after fundamental 3K or 4K redshifts, where K is Karlsson's constant: Spectra show that redshifts 3K (or 4K) exactly bring  absorbed Lyman beta (or gamma) line on  Lyman alpha: redshift almost disappears, and gas lines are intensely absorbed in the absence of alpha absorption; c) Redshifts occur in regions where light at alpha frequency is poorly absorbed due to permanent redshift, except when excitation of hydrogen to 2P level is sufficient for a superradiant flash emission at alpha frequency. This causes an intense absorption of high radiance rays from quasar, so the absorption of a line at current Lyman alpha frequency. Lightning and pumping produce relaxation oscillations that write many absorption lines. Redshifts by H atoms in 2P levels are due to parametric interactions composed of Impulsive Stimulated Raman Scatterings (ISRS):  excited hydrogen atoms catalyze energy exchanges between observed ray and background cold thermal radiation, in agreement with thermodynamics. Description of Universe becomes much simpler, but less marvelous.

\end{abstract}
Keywords:
Line:formation.
Radiative transfer.
Scattering.

98.62.Ra Intergalactic matter; quasar absorption and emission-line systems; Lyman forest

290.5910 Scattering, stimulated Raman 

190.2640 Nonlinear optics : Stimulated scattering, modulation, etc.

\section{Introduction.}
\label{Intro}
Karlsson \cite{Karlsson} and Burbidge \cite {Burbidge} used Karlsson's constant K to represent the so-called quantization of reddshifts of quasars: Most redshifts (relative variations of frequencies of spectral lines) of quasars have a value $ Z(n) = nK$, where n is an integer from the strange serie: 3, 4, 6, .... 

Note that, with the precision allowed by the calculation of K,  deducted from thousands of measures, 3K (resp. 4K) is redshift $Z(\beta,\alpha)$ (resp. $Z(\gamma,\alpha)$ which converts frequency of $\beta$ (resp. $\gamma)$ line into frequency of $\alpha$ line \cite{MB}. Rewriting Karlsson's formula:

 $Z(p, q) = pZ(\beta,\alpha)+qZ(\gamma,\alpha,)$ 	(1)

where $p$ and $q$ are non-negative integers is justified by the absence, in usual interpretation of quasars spectra, of sharp and saturated absorption lines named "redshifted Lyman $\beta$ (or $\gamma)$ lines" \cite{Petitjean,Rauch}. Calculation of such rays from redshifted rays called "Lyman $\alpha$" leads to other $\alpha$ lines. Overlay of lines is so perfect that it products no broadning of the observed line.

Karlsson's formula links astronomical redshift to properties of hydrogen atom. The purpose of this study is allowing easy comparison of results deducted from $\Lambda$-CDM to those given here by standard spectroscopy.

\section{Karlsson-Burbidge model of a quasar surrounded exclusively by hydrogen.}
\label{mod1}

\subsection{Generalities, conventions.}
\label{disc}

In Doppler frequency shifts, emitted and received frequencies verify equation which involves, as parameters, emission and reception speeds $V_{emission}$ and $V_{reception}$ :

 $\nu_{reception}=\frac{c-V_{reception}}{c-V_{emission}}*\nu_{emission}$	(2)
 
 In a first approximation we assume that frequency shifts at various light frequencies obey an equation similar to equation (2), so that a single parameter is needed to compute frequency shifts at all frequencies, for instance the ratio $f_1/f_0$ of a particular shifted frequency over initial frequency :

 $\nu_{shifted}=\frac{f_1}{f_0}\nu_{initial}$ 	(3)

If this approximation appears too bad, it may be corrected by multiplication of $\nu_{shifted}$ by a dispersion function $D(\nu_{shifted})$, generally close to one.

Using Rydberg's formula, compute redshifts which put Lyman $\nu_\beta$ and $\nu_\gamma$ frequencies of H atom to  $\nu_\alpha$ frequency:

$Z_{(\beta,\alpha)} = (\nu_\beta-\nu_\alpha)/\nu_\alpha = [(1-1/32 -(1-1/22)]/(1-1/22) ] \approx 5/27 \approx 0.1852 \approx 3*0.0617 \approx 3*K; $ (4)
 	
$Z_{(\gamma,\alpha)} = (\nu_\gamma-\nu_\alpha)/\nu_\alpha = $[(1-1/42 -(1-1/22)]/(1-1/22) ] = 1/4 = 0,25 = 4*0.0625 $\approx 4K$. (5)

Assuming that use of  3 lowest Lyman frequencies of H atom is better than use of K, and that variation of shifts at various frequencies is obtained from a particular shift as in a Doppler redshift,  Karlsson's formula becomes simpler: shifted frequencies $\nu(p,q)$ depend only on absolute frequency $\nu_0$ , three well known frequencies and two any non-negative integers $p$ and $q$:

$\nu(p,q) = (\nu_\alpha/\nu_\beta)^p * (\nu_\alpha/\nu_\gamma)^q * \nu_0$.	(6)
 
Genuine Karlsson's formula, applied, for instance, to Lyman beta line with n=3 does not generate exactly Lyman alpha frequency, while formula (6) does. We  choose formula (6) because choice of possible values of $p$ and $q$ seems more natural than extraction of $n$ from a less natural serie. Is it only aesthetics ?

\subsection{Hypothesis and observations.}

We assume that ``Lyman forest'' is built by absorption of a thermal emission of an extremely hot star by {\it pure}, relatively cold (2 000-50 000 K), low pressure, unexcited, atomic hydrogen.

Following properties of many lines of ``Lyman forest''  are directly deduced from Petitjean's paper \cite{Petitjean}:

-A- As lines are sharp, widening of lines by collisions must be negligible, pressure of gas must be very low.

-B- To obtain absorption of Lyman lines at several frequencies, a redshift process of electromagnetic waves is necessary. Here, we make only hypothesis of formula (3) about this redshift.

-C- A single, unshifted Ly$_\beta$ is observed; no unshifted Ly$_\gamma$ appears because it does not remain absorbable energy at high frequencies after redshifts of thermal emission profile. 

\medskip
Absence of an absorption line may result from :
  
-a- Absence of emitted energy around frequency of line.

-b- A permanent shift of light frequencies dilutes absorption, so that all absorbed (or emitted) lines have width of shift and lines are weak, not observable. Accordingly, absorption (or emission) of sharp lines requires a stop of frequency shifts.

-c- Accurate superposition of observed lines results from the choice of redshift equations:
Absence of shifted $\beta$ and $\gamma$ lines while sharp $\alpha$ lines are observed, is due to superposition of these lines with $\alpha$ lines.

\medskip
In  $\Lambda$-CDM  theory, sharpness of saturated absorbed lines requires contradictory conditions: 

- Column density of gas must be large for saturation;

- Absorbing gas must be thin to avoid a broadening of lines by frequency shift during absorption.

- Pressure of gas must be low to avoid collisional broadening of line.

Thus gas must be in filaments which are only detected on paths from quasars.

\medskip
Supposing that redshift is related to a physical property of gas, condition is:

- Light is redshifted except if an absorbed line is at Lyman $\alpha$ frequency. Thus redshift requires a Lyman  $\alpha$ absorption, that is generation of 2P atomic hydrogen.

\subsection{Building a spectrum by Karlsson's redshifts.}

Figure 1 represents a canvas of atomic hydrogen spectrum for building any absorption spectrum by addition of lines, in particular able to play the role of $Ly_\beta$ or $Ly_\gamma$ lines.
\begin {figure*}

\label{spectre}
\centering
\includegraphics[width=16cm]{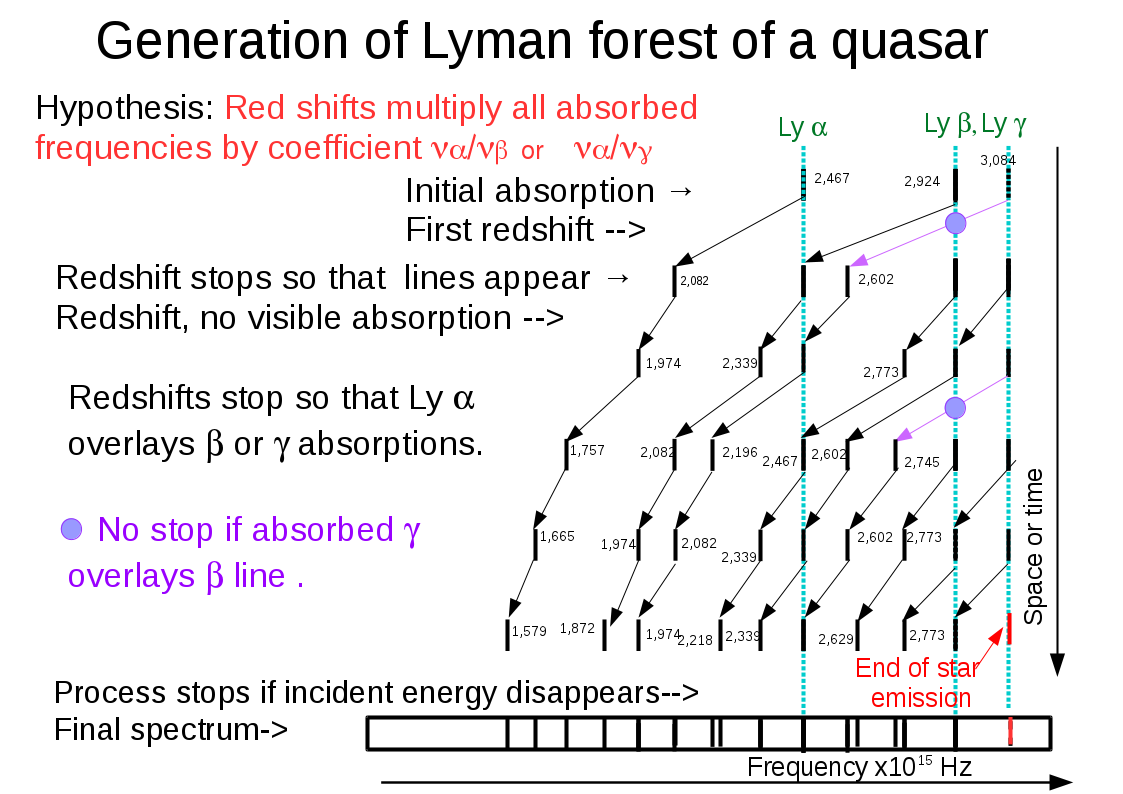}

\caption{Generation of lines of Lyman forest of a quasar by coincidence of line $\alpha$ of gas with previously absorbed, shifted $\beta$ or $\gamma$. During stop of redshift Ly$_\beta$ and Ly$_\gamma$ lines are absorbed. Lines of other local gas may be also absorbed and may later play the role of Ly$_\beta$ and Ly$_\gamma$ lines if their frequencies are larger than $\nu_\alpha$. Written frequencies do not take into account dispersion of hyperfine polarizability of H atom.}
\end{figure*}

Rules used to build spectrum taking into account only hydrogen atoms are simple :

  - i - At start, close to star, we suppose that $\alpha, \beta$ and $\gamma$ lines have been absorbed.
  
  - ii - It appears a frequency shift until an absorbed, shifted line of initial frequency $\nu$ reaches $\nu_\alpha$ frequency. Thus all absorbed frequencies have been multiplied by $\nu_\alpha/\nu$, coefficent lower than 1. In pure atomic hydrogen $\nu$ may be $\nu_\beta$ or $\nu_\gamma$, assuming that higher frequency lines are too weak.
  
  - iii - During stop of frequency shift, the three main lines of H could be absorbed, but there remain no energy at Ly$_\alpha$ frequency. 
  
   - iv - Assume that Ly$_\beta$ absorption produces a very weak redshift. During this negligible redshift, gas lines are visibly absorbed. If weak redshift is able to shift absorbed frequencies off Ly$_\alpha$ absorption line before full absorption of light at Ly$_\beta$ frequency, fast redshift restarts, go to - ii -. Else there is no more redshift, absorbed Ly$_\beta$ line is visible, but Ly$_\gamma$ line is probably not because its frequency is probably larger than the shifted high frequency limit  of emission of star. 
   
   In pure H, total redshift results from several relative shifts Z$_{\beta, \alpha}$ and  Z$_{\gamma,\alpha}$ which correspond to multiplication of light frequencies by $\nu_\alpha/\nu_\beta$ or $\nu_\alpha/\nu_\gamma$. 
   
Computed lower frequencies written on figure are not very good. They can be corrected by use of an unique dispersion function $D(\nu)$ equal to $\nu$ at and in neighborhood of $\nu_\alpha$ frequency.

Thus, formula 6 becomes:

$\nu(p,q) = D[(\nu_\alpha/\nu_\beta)^p * (\nu_\alpha/\nu_\gamma)^q * \nu_0]$. (7)

Determination of $D$ function requires either study of a good spectrum or calculation of dispersion of polarizability of H atom in state 2P.

\subsection{Introduction of other gas in spectrum building.}
Lines found in figure 1 appear in real spectra, after a correction of their frequencies by the chromatic dispersion factor  $F(\nu)$. But many lines have an other origin: various lines absorbed at frequencies larger than $\nu_\alpha$ may play the role of Ly$_\beta$ and Ly$_\gamma$, multiplying density of lines.

Figures much more complex than fig. 1 can be drawn.

\subsection{Structuring space.}
\label{struc}
Redshifts stop if frequency of an absorbed line becomes $\nu_\alpha$ and, for instance, not if absorbed $\nu_\gamma$ frequency reaches $\nu_\beta$ frequency. This shows that redshift results on generation of 2P hydrogen atoms.

We have supposed that it exists redshifts in some places. Thus, assuming that quasar is far from other stars, space is divided into spherical shells where generated 2P atoms shift light frequencies, and shells in which there is no 2P atoms and no frequency shifts.

As quasars are not alone in space, perturbation of generation of these regions may result from pumpings of atoms by light of other stars. Thus region in which quantized redshifts appear must be small enough to avoid an important lighting by other stars: This explains Burbidge's selection of quasars.

To avoid different pumpings by light coming from different regions of surface of the star, redshift must appear only at a distance much larger than size of the star, that is in a region where pressure of gas is low.

These conditions are evidently not verified for a set of relatively close stars as a galaxy. As the star must produce very high light frequencies, if it is small it must be extremely hot. An hypothesis which seems valuable is an accreting neutron star, a type of stars which were never found in nebulae while they should be bright enough to be seen.

\section{Dynamics of space structures: Absorption of Lyman $\alpha$ frequency by competition of modes.}
\label{flash}
In model described in \ref{mod1}, we obtained spherical shells of hydrogen atoms either excited by absorption of Lyman $\nu_\alpha$ frequency, or in ground state. But it is difficult to obtain the large number of lines observed for instance between Ly$_\alpha$ and  Ly$_\beta$ lines ( Ly$_\alpha$ forest ).

In \ref{struc}, we took only ito account the lack of absorption at $\nu_\alpha$ frequency to explain stops of redshifts. Let us consider a spherical shell in which Lyman alpha line is absorbed, therefore light is redshifted.

 The fraction of excited atoms grows without reaching the limit, not very lower than 1/2, which corresponds to the very high temperature of spectral radiance of star at Lyman $\alpha$ frequency. Einstein's coefficient B becomes large, so that intense superradiant emissions can burst. Flash of-excites atoms brutally and causes an intense absorption of other rays at $\alpha$ frequency by competition of modes. Thus an absorption line appears at Lyman $\alpha$ frequency. This process is in conformity with thermodynamics because spectral radiance of stellar ray is higher than that in superradiant mode. After flash, gas restarts an excitation. Process absorption-flash reproduces without any simple computation of relaxation period, therefore of position of new absorbed lines.  

It is difficult to discuss on frequency and stability of relaxation oscillations which generate these absorptions, thus on stability of spectra.  However, the medium is probably much more homogeneous and stable than the medium which provides aurora borealis. Frequencies must be adjusted
 from observation. It is also difficult to discuss geometry of flashes, seen as flares, probably induced by rays emitted by the star into directions close to direction of observation of the star.

\section{Physical interpretation.}
\label{isrs}
\subsection{Propagation of light in 1S and 2P atomic hydrogen.}
In 1S state, hydrogen has 1420 MHz  quadrupolar resonance frequency, (period T = 0.7 ns;  wavelength $\lambda$ =  21 cm).

In its first excited state, hydrogen atom has quadrupolar resonance frequencies: 178 MHz (T = 5.6 ns, $\lambda$ = 1,7 m) in state 2S$_{1/2}$, 59 MHz (T = 17 ns,  $\lambda$ = 5 m) in  2P$_{1/2}$ and 24 MHz (T = 42 ns, $\lambda$ =  12 m) in 2P$_{3/2}$.

A qualitative difference between laboratory and space ISRS is that, in labs, ISRS  is observed using femtosecond laser pulses while ordinary incoherent light is made of longer, around 1 nanosecond  pulses. For space coherence, hyperfine periods in excited states, must be longer than length of pulses (1ns) in ordinary light as required by conditions of space coherence of "Impulsive Stimulated Raman Scattering" (ISRS) : {\it Length of pulses must be shorter than all involved time constants} \cite{GLamb}.

Space coherence of incident and scattered light in ISRS allows an interference of exciting and scattered light, which produces an intermediate frequency, that is shifts frequency of incident light and preserves the geometry of light beams.
 
The other condition for ISRS is: Collisional time must be longer than 1 nanosecond, that is pressure must be very low, so that pressure broadening of lines is low.

These conditions make ISRS in space very weak compared with ISRS in labs of chemistry, which uses around  10 femtosecond laser pulses, around $k=10^5$ times shorter than light pulses of temporally incoherent light:

- Division of pressure by k divides the shift by k.

- Division of quadrupolar frequency by k divides the shift by k twice:

  - - Once mixing incident and Raman frequency shifted of hyperfine frequency;
 
  - - Once by division of difference of populations of hyperfine levels, assuming thermal equilibrium.
  
  Thus order of magnitude of path needed to observe ISRS is $\approx 10^{15}$ times longer than in laboratory experiments: astronomical paths are needed for observation.
  
\subsection{De-excitation of hyperfine levels.}
As pressure is very low, collisions between atoms are negligible. A radiative process is necessary to de-excite hyperfine levels, thus obtain a permanent Raman frequency shift.

Happily, there are cold background electromagnetic waves. Thus, the real process is not a single ISRS, but a set of ISRS such that hyperfine levels of 2P atoms have a constant excitation. Excited atoms H catalyze an exchange of energy between light beams, it is a {\it parametric interaction}. Variations of energy produce frequency shifts of light beams such that entropy is increased. This parametric interaction is usually named "Coherent Raman Effects between Incoherent Light beams" (CREIL).

\section{Main applications in astrophysics of Coherent Raman Effects on Incoherent Light.}

\subsection{Blueshift of microwaves emitted by Pioneer probes.}
Anomalous blueshifts of microwaves exchanged with Pioneer 10 and 11 probes \cite{Anderson} are explained by an "anomalous acceleration". Anomalous increase of frequency happened between 10 and 15 AU from Sun, exactly where solar wind cools into excited atomic hydrogen. It results from a transfer of energy from solar light to weak microwaves.

\subsection{Dispersion of multiplets emitted by dense regions of quasars.}
This dispersion results from dispersion of hyperfine polarizability of hydrogen atoms. There is no need of variation of fine structure constant.

\subsection{Redshifts of light by excited hydrogen.}
Excited hydrogen atoms are abundant close to hot stars and galaxies. Thus Hubble's law exaggerates distances close to these hot objects. This radial increase of distance produces, as angular distances are not changed, an increase of transverse distances. Thus, it appears anisotropic bubbles, well visible on maps of galaxies, bubbles which give a spongious appearance to these maps.

As distances of spiral galaxies are increased, their size is increased and celestial mechanics seems fail. It does not!

\section{Conclusion.}
Standard coherent spectroscopy is used to study spectra of quasars:
Karlson's formula on redshifts of quasars is demonstrated, widespread and applied to study sets of sharp, saturated absorption spectral lines. Generation of other sharp lines results from a competition of modes during flash, superradiant of-excitements of very low pressure 2P hydrogen atoms.

Interpretation of many astronomical frequency shifts is easy using coherent exchanges of energy between light beams, which increase entropy of the beams. Excited atomic hydrogen, mainly in its 3P state is a  powerful catalyst of these exchanges. it is unnecessary to introduce, as in the theory of expanding universe, new concepts to explain some observations.

An other coherent optical effect should be used in astrophysics: Superradiance shows the limbs of Strömgren spheres as possibly dotted circles. Strömgren's sphere of supernova 1987A is strangulated into an hourglass by absorption of energy by planets. Multiphotonic interactions, superradiance and competition of modes explain that the star disappeared when the beautiful rings appeared.

\medskip
Astrophysicists should learn some coherent spectroscopy or collaborate with spectroscopists.


\begin{thebibliography}{00}
\bibitem{Karlsson}Karlsson, K. G., Astron. Astrophys., 239, 50-56 (1990). 
\bibitem{Burbidge}Burbidge, G., ApJ., 154, L41-L48 (1968). 

\bibitem{MB}Moret-Bailly, J., AIP conf. proc., 822, 226-238 (2006)
\bibitem{Petitjean}Petitjean, P. , Ann. Physique, 24, 1-126 (1999).
\bibitem{Rauch}Rauch M. , Annu. Rev. Astron. Astrophys., 36, 267-316 (1998).
\bibitem{GLamb}Lamb G. L. Jr., Rev. Mod. Phys., 43, 99-124 (1971). 
\bibitem{Anderson}Anderson J. D., Lau E. L., Turyshev S. G., Laing P. A. \& Nieto M. M., Phys. Rev. Lett. 81, 2858, (1998).
\end{thebibliography}
\end{document}